\documentstyle[prl,aps,floats,epsf]{revtex}

\begin{document}
\twocolumn[\hsize\textwidth\columnwidth\hsize\csname
@twocolumnfalse\endcsname

\draft
%\title{Quark spectra, topology and random matrix theory}
\title{\vspace*{-1.2cm}\hfill\mbox{\small FSU-SCRI-99-08}\\
\hfill\mbox{\small UCD 1999-3}\\
\vspace*{0.2cm}Quark spectra, topology and random matrix theory}
\author{
Robert G. Edwards and Urs M. Heller}
\address{
SCRI, Florida State University, 
Tallahassee, FL 32306-4130, USA}
\author{ Joe Kiskis}
\address{ Dept. of Physics, University of California,
Davis, CA 95616}
\author{ Rajamani Narayanan}
\address{Dept. of Physics, Bldg. 510A,
Brookhaven National Laboratory,
P. O. Box 5000, Upton, NY 11973}

\maketitle 

\begin{abstract}
Quark spectra in QCD are linked to fundamental properties of the theory
including the identification of pions as the Goldstone bosons of spontaneously
broken chiral symmetry. The lattice Overlap-Dirac operator provides a
nonperturbative, ultraviolet-regularized description of quarks with the correct
chiral symmetry. Properties of the spectrum of this operator and their relation
to random matrix theory are studied here.
In particular, the predictions from chiral random matrix theory in
topologically non-trivial gauge field sectors are tested for the first time.
\end{abstract}
]

An important property of massless QCD is the spontaneous breaking of
chiral symmetry. The associated Goldstone pions dominate the low-energy,
finite-volume scaling behavior of the Dirac operator spectrum in the
microscopic regime, defined by $1/\Lambda_{QCD} << L << 1/m_\pi$, with
$L$ the linear extent of the system~\cite{LS}. This behavior, in turn,
can be characterized by chiral random matrix theory (RMT), which lead to
a revival of RMT, first used to understand the energy levels of nuclear
matter~\cite{porter}. What enters into the RMT description of the 
low-energy, finite-volume scaling behavior are
some symmetry properties of the Dirac operator and the sector of fixed
topological charge under consideration~\cite{SV,JV94}. The RMT predictions
are universal in the sense that only the symmetry properties, but not
the form of the potential matters ~\cite{univ}. Furthermore, the
properties tested in this letter can be derived directly from the effective,
finite-volume partition functions of QCD of Leutwyler and Smilga, without
the detour through RMT~\cite{noRMT}, though RMT nicely and succinctly
describes and classifies all these properties. The topological
charge enters the RMT prediction via the number of fermionic zero modes,
related to the topological charge through the index theorem. The symmetry
properties of the Dirac operator fall into three classes, corresponding to
the chiral orthogonal, unitary, and symplectic ensembles~\cite{JV94}.
Examples are fermions in the fundamental representation of gauge group
SU(2) for the orthogonal ensemble, fermions in the fundamental
representation of gauge group SU($N_c$) with $N_c \ge 3$ for the unitary
ensemble, and fermions in the adjoint representation of gauge group
SU($N_c$) for the symplectic ensemble.

\begin{figure*}[t]
\vspace*{-10mm} \hspace*{-0cm}
\begin{center}
\epsfxsize = 0.8\textwidth
\centerline{\epsfbox[100 230 550 390]{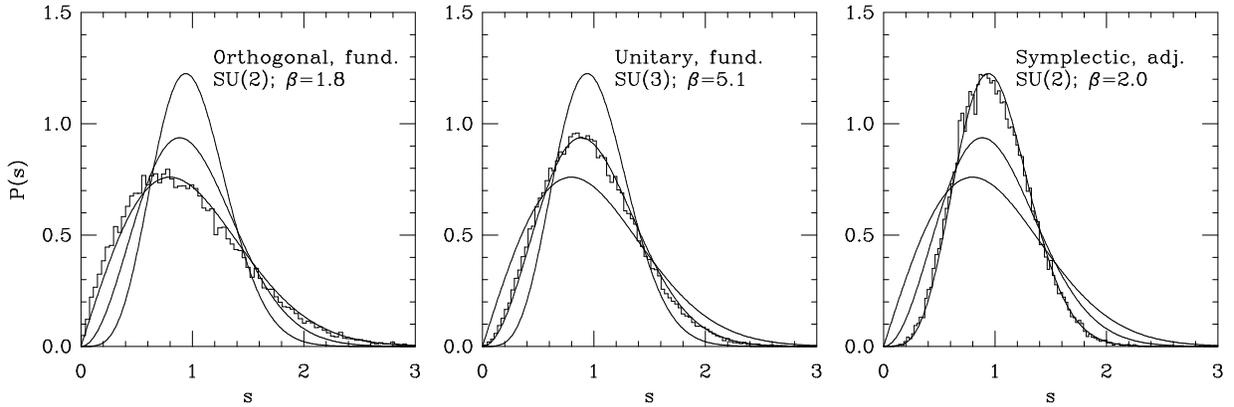}}
\end{center}
\caption{
Plots of the unfolded level spacing distribution $P(s)$ versus $s$
for the various ensembles. The three curves are the predictions
for the three ensembles where the sharpest curve is for
the symplectic ensemble, followed by the curve for the unitary ensemble
and the broadest curve is for the orthogonal ensemble.
}
\label{fig:level}
\end{figure*}

The classification according to the three different chiral random matrix
ensembles is intimately connected to the chiral symmetry properties of
the fermions. A good non-perturbative regularization of QCD should therefore
retain those chiral symmetry properties. Until recently such a regularization
was not known. The next best thing were staggered fermions, which at least
retained a reduced chiral-like symmetry on the lattice. Indeed, staggered
fermions were used to verify predictions of chiral RMT~\cite{numRMT1,numRMT2},
albeit with two important shortcomings: (i) staggered
fermions in the fundamental representation of SU(2) have the symmetry
properties of the symplectic ensemble~\cite{HV95}, not the orthogonal
ensemble as continuum fermions, while adjoint staggered fermions belong to
the orthogonal ensemble~\cite{EHNstAdj}, not the symplectic one.
(ii) staggered fermions do not have exact zero modes at finite lattice
spacing~\cite{KSzeromode}, even for topologically non-trivial
gauge field backgrounds, and thus seem to probe only the $\nu=0$ predictions
of chiral random matrix theory~\cite{numRMT1,numRMT2}.

The development of the overlap formalism for chiral fermions on the
lattice~\cite{over} recently lead to the massless Overlap-Dirac operator,
a lattice regularization for vector-like gauge theories that retains
the chiral properties of continuum fermions on the lattice~\cite{herbert}.
In particular, the continuum predictions of chiral random matrix theory
should apply. Overlap fermions have exact zero modes in topologically
non-trivial gauge field backgrounds~\cite{ehn1}, allowing, for the first
time, verification of the RMT predictions in $\nu \ne 0$ sectors.
The nice agreement we shall describe further validates the chiral RMT
predictions on the one hand and strengthens the case for the usefulness
of the Overlap regularization of massless fermions on the other hand.

The massless Overlap-Dirac operator~\cite{herbert} is given by
\begin{equation}
D={1\over 2} \left [ 1 + \gamma_5 \epsilon (H_w(m)) \right ] ~.
\label{eq:over}
\end{equation}
Here, $\gamma_5 H_w(-m)$ is the usual Wilson-Dirac operator
on the lattice and $\epsilon$ denotes the sign function. The mass $m$ has
to be chosen to be positive and well above the critical mass for Wilson
fermions but below the mass where the doublers become light on the lattice.
In this letter, we will be interested in the low lying eigenvalues of
$H=\gamma_5 D$, which is a hermitian operator.
Relevant properties of this operator can be found in Ref.~\cite{ehn1}.
We will use the Ritz algorithm~\cite{Ritz} applied to $H^2$ to obtain the
lowest few eigenvalues. The numerical algorithm involves the action of $H$ on
a vector and for this purpose one will have to use a representation of
$\epsilon(H_w(m))$. We will use the rational approximation discussed
in Ref.~\cite{ehn1,ehn2}. 

For fermions in the fundamental representation of gauge group SU(2), the
Overlap-Dirac operator is real since the underlying Wilson-Dirac operator
is real~\cite{su2}. They are therefore expected to fall into the chiral
orthogonal ensemble. For fermions in the adjoint representation the
spectrum of the Overlap-Dirac operator is doubly degenerate, since the
spectrum of the underlying Wilson-Dirac operator is doubly
degenerate~\cite{adjoint}, and they are therefore expected to belong
to the symplectic ensemble.
For three or more colors the Overlap-Dirac operator in the fundamental
representation is complex with no additional symmetries and is
therefore expected to fall into the chiral unitary ensemble.\footnote{
Fermions in U(1) abelian gauge theory are another example of the unitary
case. For numerical tests in two dimension, see \cite{Lang}.}

We will first present our results for the distribution of the ``unfolded''
level spacing~\cite{porter,unfold}. The unfolding of the eigenvalues
of $H$ are done in the following manner. Let $E_i^n$ label the non-zero
positive eigenvalues
of $H$ with $n$ labeling the configuration number and $E_i^n > E_{i-1}^n$
for all $i$. We are considering only the positive eigenvalues of $H$,
since the non-zero eigenvalues of $H$ all come in positive/negative
pairs~\cite{ehn1}. For fermions in the adjoint representation of SU($N_c$)
the spectrum of $H$ is doubly degenerate and we only keep half the spectrum,
dropping the degeneracy. Unfolding proceeds by first sorting all the
$E_i^n$ in ascending order and associating the location $N_i^n$ of $E_i^n$
in the sorted list with $E_i^n$. $N_i^n$ is referred to as the unfolded
spectrum and the level spacing is simply given by $(N_{i+1}^n-N_i^n)/N$
where $N$ is the number of configurations.
The distributions of the unfolded level spacing, $s$, in RMT
are well approximated by the various Wigner distributions~\cite{HV95}
\begin{equation}
P(s) = \cases {{\pi\over 2} s {\rm e}^{-{\pi\over 4}s^2} & orthogonal
 ensemble \cr
{32\over\pi^2} s^2 {\rm e}^{-{4\over\pi}s^2} & unitary ensemble \cr
{262144\over 729\pi^3} s^4 {\rm e}^{-{64\over 9\pi} s^2} & symplectic
 ensemble ~. \cr }
\label{eq:level_space}
\end{equation}

\begin{figure*}[t]
\vspace*{-10mm} \hspace*{-0cm}
\begin{center}
\epsfxsize = 0.8\textwidth
\centerline{\epsfbox[100 175 550 500]{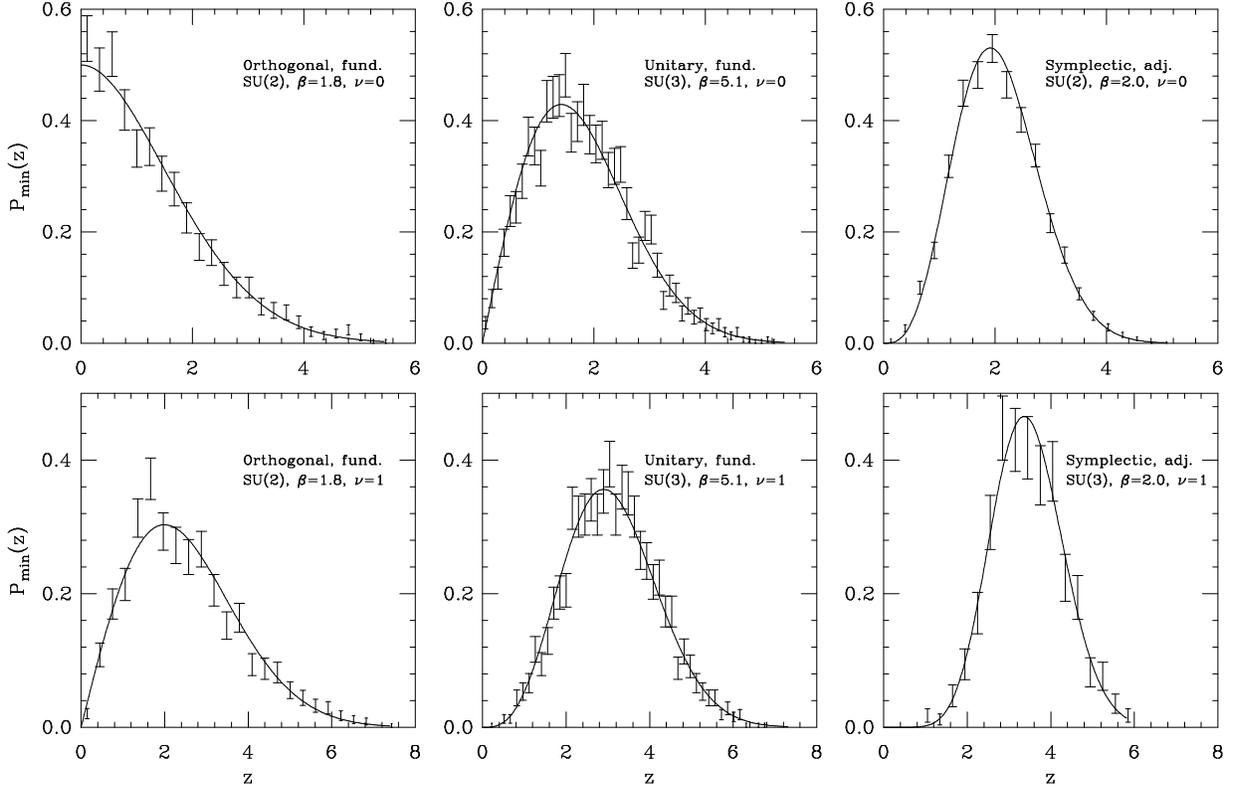}}
\end{center}
\caption{
Plots of $P_{\rm min}(z)$ versus $z$
for the various ensembles in the
lowest two topological sectors. The curve in each plot is 
a fit to the prediction
from random matrix theory with the best value for the chiral condensate.
}
\label{fig:pmin}
\end{figure*}

In our numerical simulations we computed the low lying spectrum of the
Overlap-Dirac operator in the fundamental representation on pure gauge
SU(2) configurations with $\beta=1.8$ as an example of the chiral orthogonal
ensemble, on pure gauge SU(3) configurations with $\beta=5.1$ as an example
of the chiral unitary ensemble, and in the adjoint representation on
pure gauge SU(2) configurations with $\beta=2.0$. The lattice size was
$4^4$ in all cases.
The various level spacing distributions are shown in Fig.~\ref{fig:level}.
There is very clear evidence that the SU(2) and SU(3) ensembles
with fermions in the fundamental representation fall into the orthogonal
and unitary ensemble, respectively, and the SU(2) ensemble with fermions
in the adjoint representation falls into the symplectic ensemble.

We next turn to the distribution of the lowest eigenvalue for the various
ensembles. 
Chiral RMT predicts that these distributions are universal
when they are classified according to the three ensembles and according to the
number of exact zero modes $\nu$ within each ensemble and then considered
as functions of the rescaled variable $z=\Sigma V \lambda_{\rm min}$.
Here $V$ is the volume and $\Sigma$ is the infinite volume value of the
chiral condensate $\langle \bar \psi \psi \rangle$ 
determined up to an overall wave function normalization, which is dependent 
in part on the Wilson--Dirac mass $m$.
RMT gives for the distribution of the rescaled lowest eigenvalue
for the orthogonal ensemble, expected to apply to the fermions in the
fundamental representation of SU(2), in the $\nu=0$ and $\nu=1$
sector~\cite{Forrester}
\begin{equation}
P_{\rm min}(z) = \cases { 
{2+z\over 4} {\rm e}^{-{z\over 2}-{z^2\over 8}} & if $\nu=0$ \cr
{z\over 4} {\rm e}^{-{z^2\over 8}} & if $\nu=1$ ~. \cr } 
\label{eq:Pmin_OE}
\end{equation}
For the unitary ensemble, expected to apply to the fermions in the
fundamental representation of SU($N_c$) with $N_c \ge 3$, the RMT
predictions are~\cite{Forrester,PHD}
\begin{equation}
P_{\rm min}(z) = \cases { 
{z\over 2} {\rm e}^{-{z^2\over 4}} & if $\nu=0$ \cr
{z\over 2} I_2(z) {\rm e}^{-{z^2\over 4}} & if $\nu=1$ \cr
{z\over 2} \left[ I_2^2(z) - I_1(z) I_3(z) \right] {\rm e}^{-{z^2\over 4}}
 & if $\nu=2$ ~. \cr}
\label{eq:Pmin_UE}
\end{equation}
Finally, for the symplectic ensemble, expected to apply to the fermions
in the adjoint representation, the RMT prediction
is~\cite{Forrester,Kaneko,numRMT2}
\begin{equation}
P_{\rm min}(z) = \cases {
\sqrt{\pi\over 2} z^{3/2} I_{3/2}(z) e^{-{z^2\over 2}} & if $\nu=0$ \cr
{2\over {(2\nu+1)! (2\nu+3)!}} z^{4\nu+3} {\rm e}^{-z^2\over 2}
T_\nu(z^2) & if $\nu > 0$ ~. \cr}
\label{eq:Pmin_SE}
\end{equation}
We are interested in $\nu=1$ since the eigenvalues are doubly degenerate.
A closed form expression is not known for $T_\nu(z^2)$. Instead, a 
rapidly converging series~\cite{Kaneko,numRMT2} based on 
partitions of integers is available, namely
$T_\nu(x) = 1 + \sum_{d=1}^{\infty} a_d x^d$
where
\begin{eqnarray}
\lefteqn{a_d = \sum_{{|\kappa|=d}\atop{l(\kappa)\le2\nu+1}} \prod_{(i,j)} 
  {{(2\nu+2j-i)}\over{(2\nu+2j-i+4)}}\times} \nonumber \\ 
 && {1\over{(\kappa_j'-i+2(\kappa_i-j)+1) (\kappa_j'-i+2(\kappa_i-j)+2)}} \quad .
\end{eqnarray}
Here, the integer partition $\kappa=\{\kappa_1,\kappa_2,\ldots,\kappa_d\}$ has
length $d$ and weight $l(\kappa)=\sum_i \kappa_i$. A pair of integers is
associated with $\kappa=\{(i,j)| 1\le i\le l(\kappa), 1\le j\le
\kappa_i\}$, and $\kappa_i' = {\rm Card}(j| \kappa_j \ge i)$ is the conjugate
partition.

We compare the RMT predictions with our data in Fig.~\ref{fig:pmin}.
If one knows the value of the chiral condensate in the infinite volume
limit, $\Sigma$, the RMT predictions for $P_{\rm min}(z)$ are
parameter free. On the rather small systems that we considered here,
we did not obtain direct estimates of $\Sigma$. Instead, we made
one-parameter fits of the measured distributions, obtained from
histograms with jackknife errors, to the RMT predictions,
with $\Sigma$ the free parameter. Our results, together with some
additional information about the ensembles, are given in
Table~\ref{tab:Sigma}. We note the consistency of the values for $\Sigma$
obtained in the $\nu=0$ and $\nu=1$ sectors of each ensemble.
Alternatively, we could have used the value of $\Sigma$ obtained in
the $\nu=0$ sector, to obtain a parameter free prediction for the
distribution of the rescaled lowest eigenvalue in the $\nu=1$
sector. Obviously, the predictions would have come out very well.

For the two ensembles with the fermions in the fundamental representation,
we also found 81 (for SU(2)), and 147 (for SU(3)) configurations with
two zero modes and 1 and 3 with three zero modes. For the orthogonal
ensemble, we are not aware of a prediction for $P_{\rm min}(z)$ in the
$\nu=2$ sector, while for the unitary ensemble our data, albeit with
very limited statistics, agrees reasonable well with the parameter free
prediction, eq.~(\ref{eq:Pmin_UE}) with $\Sigma$ from Table~\ref{tab:Sigma}.

For fermions in the adjoint representation, we keep only one of each
pair of doubly degenerate eigenvalues, so $\nu=1$ is the sector where
there are two exact zero modes. Such gauge field configurations cannot
be assigned an integer topological charge since integer charges
give rise to zero modes in multiples of four~\cite{ad_index}, and we
note there are a significant number of configurations with two
zero modes as seen in Table~\ref{tab:Sigma}. The good agreement
with the RMT prediction found in this case lends further support to
the existence of configurations with fractional topological
charge~\cite{adjoint}.

In this letter we have tested the predictions of chiral random
matrix theory using the Overlap-Dirac operator on pure lattice gauge
field ensembles. We find full agreement with the
unfolded level spacing distributions on all three ensembles.
We also found that the distribution of the lowest eigenvalue in the
different topological sectors fitted well with the predictions of
chiral RMT, with compatible values for the chiral condensate from the
different topological sectors. This is the first test of the influence
of topology on the Dirac spectrum in the microscopic regime.

\begin{table}
\caption{The chiral condensate, $\Sigma$, from fits of the distribution
of the lowest eigenvalue to the RMT predictions. The third column gives
the Wilson-Dirac mass parameter used, the fourth the number of
configurations, $N_\nu$, in each topological sector and the last
the confidence level of the fit.}
\label{tab:Sigma}
\begin{tabular}{|l|c|c|c|r|c|c|}
\hline
 Repr. & $\beta$ & $m$ & $\nu$ & $N_\nu$ & $\Sigma$ & $Q$ \\
\hline
 SU(2) fund. & 1.8 & 2.3 & 0 & 1293 & 0.2181(51) & 0.915 \\
 SU(2) fund. & 1.8 & 2.3 & 1 & 1125 & 0.2155(37) & 0.205 \\
\hline
 SU(3) fund. & 5.1 & 2.0 & 0 & 2714 & 0.0827(8)~ & 0.444 \\
 SU(3) fund. & 5.1 & 2.0 & 1 & 2136 & 0.0829(6)~ & 0.656 \\
\hline
 SU(2) adj.  & 2.0 & 2.3 & 0 & 1251 & 0.2900(30) & 0.611 \\
 SU(2) adj.  & 2.0 & 2.3 & 1 &  254 & 0.2931(45) & 0.313 \\
\hline
\end{tabular}
\end{table}

\acknowledgements
This research was supported by DOE contracts 
DE-FG05-85ER250000 and DE-FG05-96ER40979.
Computations were performed on the workstation cluster
at SCRI and the Xolas computing
cluster at MIT's Laboratory for Computing Science.
We thank P.H. Damgaard for discussions.

\end{document}